\documentclass[twocolumn]{aastex6}
\usepackage{amsmath}

\usepackage{hyperref}
\hypersetup{
	pdftitle={Internal Gravity Waves in the Magnetized Solar Atmosphere. I. Magnetic Field Effects},
	pdfauthor={Vigeesh}, 
	pdfsubject={The Astrophysical Journal - The Sun and the Heliosphere},   
	pdfkeywords={magnetohydrodynamics (MHD), Sun: atmosphere, Sun: granulation, Sun: magnetic fields, Sun: photosphere, waves, internal gravity waves}
}

\shorttitle{Internal gravity waves in the magnetized solar
atmosphere}
\shortauthors{Vigeesh et al.}

\begin{document}

\title{Internal Gravity Waves in the Magnetized Solar Atmosphere.\\
I. Magnetic field effects}

\author{{G.~Vigeesh}\altaffilmark{1,}\altaffilmark{2}, {J.~Jackiewicz}\altaffilmark{2}, 
and {O.~Steiner}\altaffilmark{1,3}}

\affil{\altaffilmark{1}Kiepenheuer-Institut f\"{u}r Sonnenphysik, Sch\"{o}neckstrasse
6, 79104 Freiburg, Germany}

\affil{\altaffilmark{2}New Mexico State University, Department of Astronomy, P.O. Box
30001, MSC 4500, Las Cruces, NM 88003, USA}

\affil{\altaffilmark{3}Istituto Ricerche Solari Locarno (IRSOL), via Patocchi
57--Prato Pernice, 6605 Locarno-Monti, Switzerland}

\email{vigeesh@leibniz-kis.de} 

\begin{abstract}
Observations of the solar atmosphere show that internal gravity waves are
generated by overshooting convection, but are suppressed at locations of
magnetic flux, which is thought to be the result of mode conversion into
magneto-acoustic waves. Here, we present a study of the acoustic-gravity wave
spectrum emerging from a realistic, self-consistent simulation of solar
(magneto-)convection. A magnetic field free, hydrodynamic simulation and a
magneto-hydrodynamic (MHD) simulation with an initial, vertical, homogeneous
field of 50\,G flux density were carried out and compared with each other to
highlight the effect of magnetic fields on the internal gravity wave propagation
in the Sun's atmosphere.  We find that the internal gravity waves are absent or
partially reflected back into the lower layers in the presence of magnetic
fields and argue that the suppression is due to the coupling of internal gravity
waves to slow magneto acoustic waves still within the high-$\beta$ region of
the upper photosphere. The conversion to Alfv\'{e}n waves is highly unlikely in
our model because there is no strongly inclined magnetic field present. We argue
that the suppression of internal waves observed within magnetic
flux-concentrations may also be due to non-linear breaking of internal waves due
to vortex flows that are ubiquitously present in the upper photosphere and the
chromosphere.
\end{abstract}


\section{Introduction}{\label{s:introduction}}
The solar atmosphere provides a favourable environment for the generation and
propagation of internal gravity waves (or internal waves). Turbulent convection
from subsurface regions penetrating locally into a stably stratified medium
above it, is thought to excite internal waves, along with acoustic waves. These
waves couple the lower atmosphere with the higher layers by transporting energy,
and presumably contributing to the heating of the upper solar atmosphere.
However, the short radiative timescales and the presence of strong magnetic
fields in these regions influence the internal waves. The effects that magnetic
fields may have on the generation and propagation of these waves are still
unknown.

Internal waves are a natural response of a gravitationally stratified medium to
any disturbance of its equilibrium state, with buoyancy acting as the
equilibrium restoring force. Internal waves are ubiquitous in the Earth's
atmosphere and have been extensively studied for their role in the circulation
patterns in the oceans and the terrestrial atmosphere. They form an essential
component in the general circulation models (GCM) that provide accurate global
weather predictions. The downward propagating, east-west oscillatory patterns
known as Quasi-Biennial Oscillations (QBO) observed in the Earth's atmosphere
below 35\,km in tropical latitudes are due to momentum transport by internal
waves. Tsunamis in open oceans excite internal waves that propagate up to
ionospheric heights causing traveling ionospheric disturbances
\citep{2005GeoJI.160..840A}.

Studies of internal waves in the solar atmosphere began with
\citet{1963ApJ...137..914W} 
following a suggestion by 
\citet{1960CaJPh..38.1441H}, 
a pioneer in the field of terrestrial atmospheric physics, that internal waves
could play an important role in coronal heating. Later work invoked internal
waves to explain the then elusive 5-min oscillations of the solar atmosphere
\citep{1960IAUS...12..321L,
1962ApJ...135..474L}.
The theoretical framework  put forth by 
\citet{1964ApJ...139...48M} 
tried to explain these oscillations due to frequencies below the acoustic
cut-off value, a regime where the internal gravity waves exist. Later, a number
of works explored the existence of trapped internal gravity waves due to a
temperature dip
\citep{1965ApJ...142..335U,
1967ApJ...147..181U} 
or due to ionization effects 
\citep{1971SoPh...16...51T} 
and related those to the observed solar oscillations. These studies later gave
way to trapped acoustic waves in the solar interior as the sole agent
responsible for the oscillations
\citep{1970ApJ...162..993U,
1971ApL.....7..191L}.
Despite the fact that they did not play a role in the observed oscillations,
studies of internal waves continued in view of explaining the heating of the
upper atmosphere.

\citet{1967IAUS...28..429L} 
suggested that internal waves are efficiently generated by ``tongues of
turbulence'' that reach up into the photosphere where they contribute to
atmospheric heating.
\citet{1967SoPh....2..385S}
discussed the generation of internal waves by turbulence in an isothermal,
stratified atmosphere. However, the short radiative relaxation times in the
photosphere raised questions about the mere existence of internal waves in these
regions
\citep{1967ARA&A...5...67S,
1969SSRv....9..713K,
1970A&A.....4..189S,
1973SoPh...30..319C,
1980SSRv...27..301L}.
Analytical studies of the complete magneto-acoustic-gravity (MAG) spectrum in a
simple stratified atmosphere have been carried out by a number of authors
starting with
\citet{1958ApJ...127..459F},
\citet{1982A&A...112...16Z},
\citet{1982A&A...112...84L},
\citet{1984A&A...132...45Z}, 
\citet{1992ApJ...396..311H},
\citet{1997RSPSA.453..943B},
\citet{2001ApJ...548..473C},
\citet{2015GApFD.109..168C},
to cite a few.

Some of the first observational evidences suggesting the existence of internal
waves in the solar atmosphere were presented by
\citet{1976SoPh...47..435S}
and
\citet{1978A&A....70..345C}.
An extensive study of internal waves in the solar atmosphere focusing on the
energy dissipation and their possible signatures on spectral lines was carried
out by
\citet{1981ApJ...249..349M,
1982ApJ...263..386M}.
They concluded that the energy dissipation of internal waves due to non-linear
wave-breaking is dominant in the mid-chromosphere and that they deposit all of
their energy at these heights, hardly ever reaching the corona. While the
detection of internal waves in the solar atmosphere has been questioned
\citep{1968ApJ...152..557F,
1979ApJ...231..570L},
a series of observations reported evidence of internal waves in the solar
atmosphere 
\citep{1981A&A....95..221D,
1984MmSAI..55..147S,
1987A&A...175..263S,
1989A&A...213..423D,
1991A&A...242..271M,
1991A&A...244..492B,
1991A&A...252..827K,
1993A&A...274..584K,
1997A&A...324..704S,
2001A&A...379.1052K,
2003A&A...407..735R}.

Using high spatial and temporal resolution spectroscopic observations in
multiple lines with ground and space-based telescopes and with the help of 3D
numerical simulation,
\citet{2008ApJ...681L.125S} 
reported the first ``unambiguous'' detection of propagating internal waves in a
magnetically quiet region of the solar atmosphere. They claimed that the energy
flux of internal waves was sufficient for balancing the radiative losses of the
chromosphere. They also observed that internal waves are suppressed in strong
magnetic field regions as a result of reflection and conversion to other wave
modes. Soon after,
\citet{2008MNRAS.390L..83S} 
found signatures of internal waves in temperature fluctuations derived from the
Fe\,{\textsc i} ($\lambda$\textnormal{=}532.418\,nm) spectral line, a
temperature sensitive line formed at photospheric heights, raising questions
about their presence at these heights despite strong radiative damping.
\citet{2011A&A...532A.111K} 
have reported the presence of internal waves and estimated their energy flux
using observations in the lines of Fe\,{\textsc i}
($\lambda$\textnormal{=}557.6\,nm, 543.4\,nm) that form at an average height of
380\,km and 570\,km, respectively. Recent work by
\citet{2014SoPh..289.3457N} 
also shows signature of internal waves in the SDO/HMI Dopplergrams. However, the
numerical models in their work fail to show a clear signature of internal waves.
This discrepancy may be due to the extent of the simulated domain, or the
radiative damping in the model, or the upper boundary conditions. Despite recent
observational confirmation of the existence of internal waves in the solar
atmosphere, not much research was done towards understanding the
power suppression of these waves in magnetic field regions.

Many different wave co-exist and interact with each other in the solar
atmosphere. The surface-gravity waves ($f$-mode) and the evanescent tails of the
solar $p$-modes exist in the atmosphere. In magnetic flux tubes, 
magneto-acoustic waves are generated as a result of continuous buffeting by
granules
\citep{1999ApJ...519..899H} 
and by strong inter-granular downdrafts
\citep{2011ApJ...730L..24K},
which propagate upwards and partially escape the flux tube to propagate as
acoustic waves in the medium outside
\citep{2009A&A...508..951V}. 
The magneto-acoustic waves that propagate up along the flux tubes undergo
transmission and conversion at the equipartition level, the height where the
ratio of sound speed ($c_{S}$) to Alfv\'{e}n speed ($v_{A}$) drops below 1. The
resulting fast magneto-acoustic waves get partially refracted travelling
downwards in the atmosphere and partially convert to Alfv\'{e}n waves near the
apex of the refractive wave path
\citep{2012ApJ...746...68K}.
Internal waves can also couple to magneto-acoustic and Alfv\'{e}n waves as shown
by
\citet{2010MNRAS.402..386N,
2011MNRAS.417.1162N}.
The whole sequence of wave production and coupling, starting from the solar
surface up to heights where  Alfv\'{e}n waves are produced, has to be clearly
understood in order to account for the energy distribution among various wave
modes at different heights. Radiative damping in the low-photosphere and
non-linear effects leading to wave-breaking above the mid-chromosphere,
spatially restrict the propagation of internal waves in the Sun's atmosphere,
making their observation difficult.

In this paper, we use realistic numerical simulations of the solar atmosphere to
study the acoustic-gravity wave spectrum's properties in the presence of
magnetic fields. This work is a substantial extension to the linear analysis
that was carried out by
\citet{1981ApJ...249..349M,
1982ApJ...263..386M},
that also neglected the effects of magnetic field. Realistic simulations that
take into account essential physics like non-local radiative transfer and an
equation of state that adequately describes the solar plasma are needed to
explain the observed properties of internal waves in the solar atmosphere.
Theoretical work on MAG waves has been carried out by a number of authors, but
atmospheric internal gravity waves in the presence of spatially intermittent and
temporally evolving magnetic fields is a less explored field. Whether the
presence of a magnetic field modifies the background properties and indirectly
affects the propagation of internal waves or whether the changes in the plasma
$\beta$ and magnetic field orientation restrict the occurrence of internal waves
to an even smaller region or perhaps suppresses them completely is still not
clear. This paper addresses some of these aspects with state-of-the-art
numerical simulations and attempts to fill some gaps in our understanding of
atmospheric internal gravity waves.

The paper is structured as follows: In Section~\ref{s:num_sim}, we discuss the
numerical setup, the construction of the model, and give a detailed description
of the properties of the non-magnetic and magnetic model in the context of
internal waves. In Section~\ref{s:spectral_analysis}, we carry out a spectral
analysis of the 3D simulation, where the emergent phase and energy flux spectra
are presented, highlighting the differences between the two models. In
Section~\ref{s:discussion}, we present a detailed discussion on the various
effects that can explain the differences between the two models. The summary and
conclusion of the paper is provided in Section~\ref{s:conclusion}.\\

\section{Numerical models}{\label{s:num_sim}}
The numerical simulations of solar convection presented in this paper were
carried out using the {CO$^{\rm 5}$BOLD} code
\citep{2012JCoPh.231..919F}. 
The code solves the equations of (magneto-)hydrodynamics for a fully
compressible gas with a realistic equation of state, taking non-local radiative
transfer into account. Here, we use five opacity groups, adapted from the MARCS
stellar atmosphere package
\citep{2008A&A...486..951G}.
We take a 3D snapshot from an earlier model of relaxed convection, computed
using {CO$^{\rm 5}$BOLD}, and extend the domain by tiling it in the horizontal
directions. The new computational domain has a size of
{38.4\,Mm}\,$\times$\,{38.4\,Mm}\,$\times$\,{2.8\,Mm}, with a horizontal cell
size of 80\,km and a vertical cell size varying from 50\,km in the lower part of
the computational domain down to 20\,km in the upper atmosphere, discretized on
480\,$\times$\,480\,$\times$\,120 grid cells. The domain reaches $\sim$1.5\,Mm
below the level of average Rosseland optical depth $\tau_{R}\textnormal{=}1$ 
(where we define the $z$ axis such that $\langle z(\tau_{R}\textnormal{=}1)\rangle=0$) and $\sim$1.3\,Mm above it. A constant gravity of g=275\,m\,s$^{-2}$
acts in the box. The tiling results in a periodic pattern due to the previous
periodic boundary condition. This pattern is eliminated by superimposing a random velocity
pattern, with rms value of 0.5v$_{x,y}$ (v$_{x}$ and v$_{y}$ are the horizontal components of the velocity), on the model between $z$=$-100$ and 0\,km
(below the average $\tau_{R}\textnormal{=}1$ surface) over the entire
horizontal scale and advancing the solution over several turnover timescale
(approx.\, 190\,min). Taking this solution as the initial model, a hydrodynamic
(HD) and a magneto-hydrodynamic (MHD) simulation run is carried out. For the
entire HD run, starting with the small domain, we use the Roe solver with
VanLeer reconstruction
\citep[see][for the details
on the computational methods]{2012JCoPh.231..919F}.
The HLL-MHD solver with PP reconstruction is used for the MHD run
\citep{2013MSAIS..24..100S}. 
For creating the MHD model, the extended initial HD model is embedded with a
uniform vertical field of 50\,G in the entire domain and advanced over a
magnetic field redistribution timescale of approximately 600\,s. During this
time the uniformly distributed fields are swept towards the inter-granular lanes
by granular flow, forming localised flux concentrations with magnetic field
strengths surpassing 1.5\,kG at z=0\,km. This model serves as a representation
of an internetwork region of the quiet-Sun. The HD solution is advanced for the
same duration to match with that of the MHD run. Both the hydrodynamic
(``non-magnetic'') and magneto-hydrodynamic (``magnetic'') solutions are then
advanced for 8 hours physical time with snapshots taken every 30 seconds. A
summary of the numerical setup and physical properties of the two simulated
models are shown in Table~\ref{tab:model_summary}.

\begin{table}[h!]
	\centering
	\caption{Numerical setup and physical properties of the two
		simulated models.}\label{tab:model_summary}
	\begin{tabular}{lcc}
		\hline
		\hline
		& Non-magnetic & Magnetic \\
		\hline
		Snapshot cadence & \multicolumn{2}{c}{30\,s}\\
		Duration of simulation & \multicolumn{2}{c}{8\,hrs}\\
		Computational grid & \multicolumn{2}{c}{480$\times$480$\times$120}\\
		Domain size &
		\multicolumn{2}{c}{38.4$\times$38.4$\times$2.8\,Mm$^{3}$}\\
		Computational cell size &
		\multicolumn{2}{c}{80$\times$80$\times$(50-20)\footnote{The vertical
				cell size varies from
				50\,km in the lower part of the computational domain
				down to 20\,km in the upper atmosphere.} km$^{3}$}\\
		Numerical scheme & Roe & HLL-MHD \\
		Reconstruction & VanLeer & PP/VanLeer\\
		Temperature, $T_{\rm eff}$ & 5798$\pm$3\,{\rm K} &
		5773$\pm$4\,{\rm K}\\
		Intensity contrast, $\delta I_{\rm rms}$
		&15.57$\pm$0.13\,\% &15.32$\pm$0.11\,\%\\
		\hline
	\end{tabular}
\end{table}

Periodic boundary conditions are used for the side boundaries in both models.
The velocity field, radiation, and the magnetic field components are periodic in the
lateral directions, which results in the inhibition of waves with horizontal
wavelengths larger than the width of the box. The top boundary is open for fluid
flow and outward radiation, with the density decreasing exponentially 
in the boundary cells outside
the domain. The vertical component of the magnetic field is constant across the
boundary and the transverse component drops to zero at the boundary. In both
models, the bottom boundary is set up in such a way that the in-flowing material
carries a constant specific entropy of ${\rm 1.773\times10^{9}\,erg\,g^{-1}\,K^{-1}}$ 
resulting in a radiative flux corresponding to an effective temperature 
($T_{\rm eff}$) of $\sim$5770\,K. The bottom boundary conditions for the magnetic 
fields are the same as for the top boundary.

\begin{figure}
	\centering
	\includegraphics[width=\columnwidth]{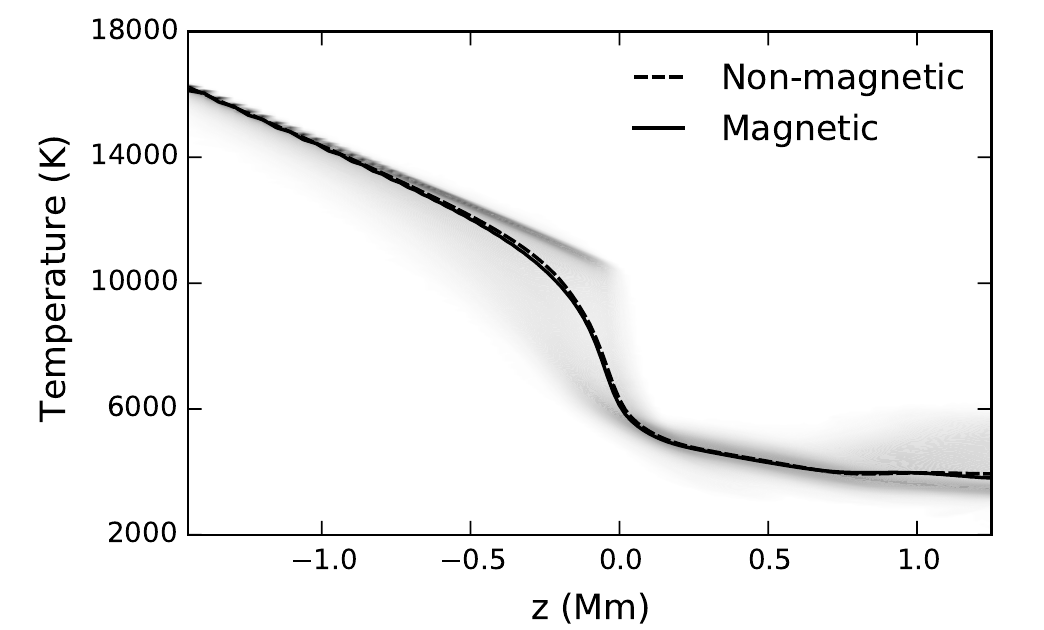}
	\caption{
		Average temperature as a function of height in the non-magnetic
		(dashed) and the magnetic (solid) model. The gray background shows the
		temperature distribution for a single snapshot of the non-magnetic run
		taken at t=4h.}
	\label{fig:temperature}
\end{figure}

\begin{figure}
	\centering
	\includegraphics[width=\columnwidth]{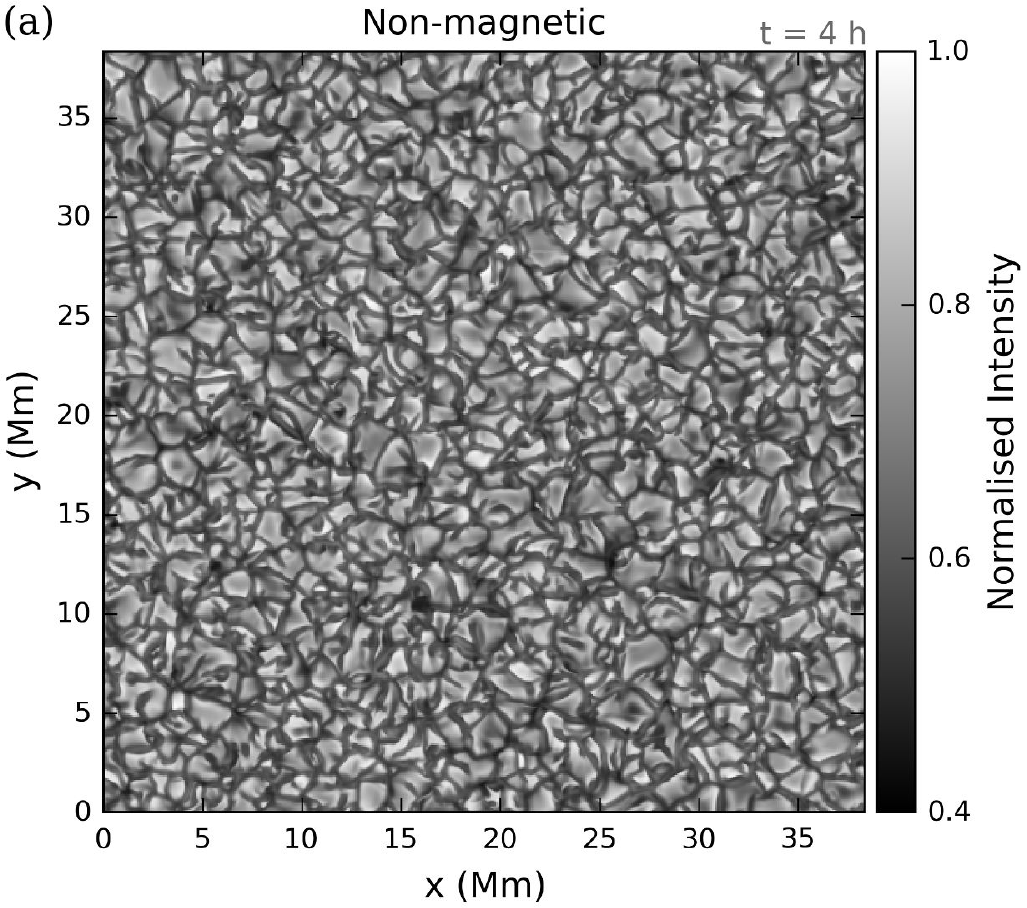}\\
	\includegraphics[width=\columnwidth]{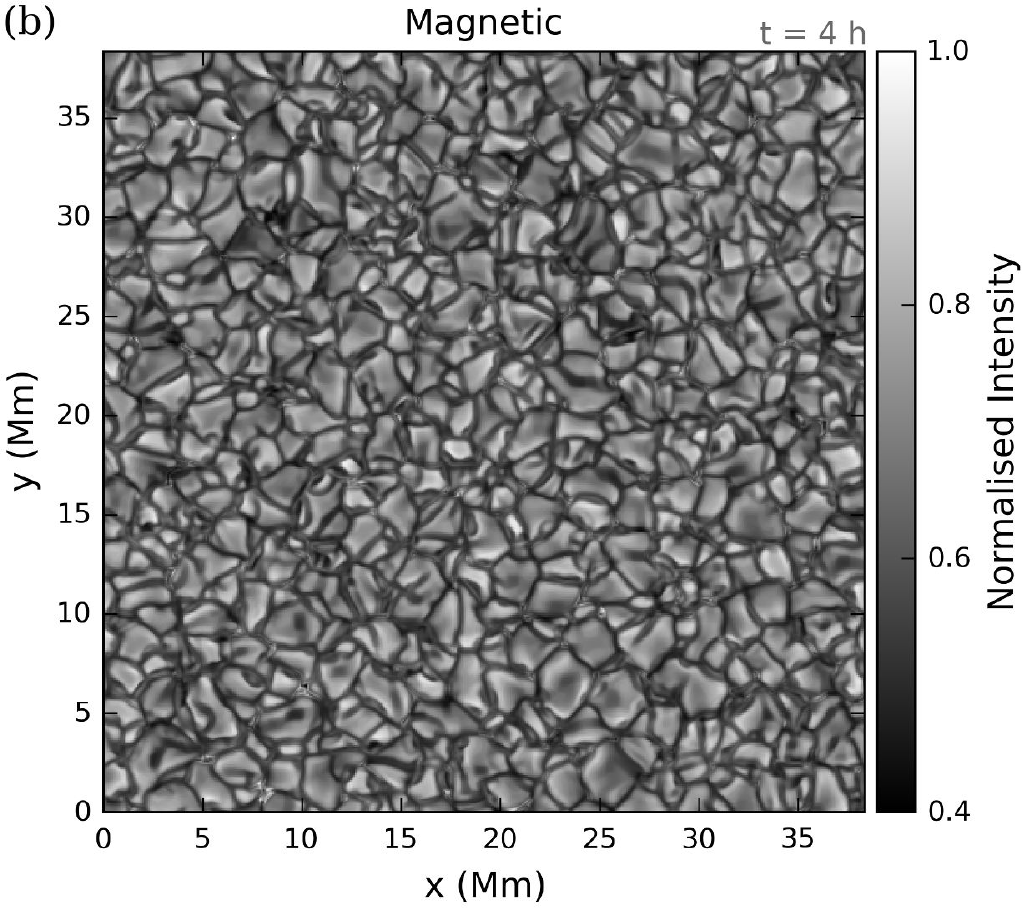}\\
	\caption{
		Emergent bolometric intensity from: a) the non-magnetic and b)
		the magnetic model at t=4h.}
	\label{fig:bolometric_intensity}
\end{figure}

The spatially and temporally averaged temperature profile of the two models is
shown in Figure~\ref{fig:temperature}. Also shown in the background is the
temperature distribution from a single snapshot of the non-magnetic model taken
at t=4h after the start of the simulation. Although the average temperature in
the upper layers becomes constant, there are instances when the temperature
increases locally, hinting to a weak shock-heated chromosphere. The two models
show exactly the same temperature profile, but the granular sizes show slight
differences.
In Figure~\ref{fig:bolometric_intensity}, we show the emergent bolometric
intensity from the two models 4~hours after the start of the simulation. It is
to be noted that, while the average size of granules in the non-magnetic model
peaks at 2\,Mm, the average granules in the magnetic model are larger. This is
due to the more diffusive nature of the HLL-MHD numerical solver, compared to
the Roe solver. However, this difference between the non-magnetic and magnetic
model does not seem to influence the overall spectra of the generated internal
gravity waves as will be further explained in Sect.~\ref{s:conclusion}. The average rms bolometric intensity
contrast, $\delta I_{\rm rms}$, of the non-magnetic and magnetic models, are
15.57\,\% and 15.31\,\%, respectively (see Table \ref{tab:model_summary}).

The large spatial and temporal coverage of the two models give us the
opportunity to study the different wave phenomena in Fourier space. All the
physical variables are decomposed into their Fourier components along the
horizontal directions and in time. In the following, we present the properties
of the model in frequency-space for a better understanding of the different wave
phenomena present in the simulation. The rest of the paper is based on this
decomposition and hence we attempt a detailed presentation.

\subsection{The dispersion relation ($k_{h}\textnormal{-}\omega$
diagram)}\label{ss:kwdiagram} 
In an infinite, homogenous, compressible medium in the absence of an external
force field, any small-amplitude perturbation propagates as acoustic wave owing
only to the compressibility of the medium. The propagation is isotropic and
non-dispersive with all the frequencies travelling	at the characteristic sound
speed ($c_{\rm s}$) in all directions. In the presence of an external force like
gravity, the propagation becomes anisotropic and acoustic waves are modified,
with waves below a certain frequency becoming vertically non-propagative.
Acoustic waves propagating horizontally, also called Lamb waves, are unaffected
and therefore are non-dispersive. A continuously stratified fluid supplies a
restoring force, in the form of buoyancy, resulting in the propagation of
internal gravity waves. The coupling of the two waves in a compressible
stratified medium, like that of the solar atmosphere, results in their
separation into gravity-modified acoustic and compressibility-modified gravity
waves. The two types of waves occupy distinct branches in the frequency-wave
number domain ($k_{h}\textnormal{-}\omega$ space) with a band of evanescent
disturbance, separating the two branches. While, stratification results in a
cut-off frequency for the acoustic waves, the effect of compressibility 
modifies the internal wave spectra at small horizontal 
wavenumbers ($k_{h}$\textless 1/(2$H_{\varrho})$, where $H_{\varrho}$ is 
the density scale height) from propagating. A detailed exposition on these 
waves is provided by
\citet{2001wafl.book.....L}.

The addition of magnetic fields to such a medium introduces waves due to the
magnetic tension and pressure forces, that couple to the other waves already
present in the medium, resulting in a spectrum of magneto-acoustic-gravity
waves. Linearizing the full MHD equations about a uniformly stratified
background state and assuming a wave-like solution, one obtains the dispersion
relation for the magneto-acoustic gravity waves. Further assuming that the presence of a
magnetic field just modifies the background atmosphere, the coupling to the
magnetohydrodynamic waves can be neglected. The dispersion relation of the waves
then reduce to
\citep[see][for a derivation]{2014masu.book.....P}

\begin{equation}
k_{z}^2 = \frac{(\omega^2 - \omega_{\rm ac}^2)}{c_{\rm s}^2} -
\frac{(\omega^2 - N^2) k_{h}^2}{\omega^2 },
\label{eq:disp_relation}
\end{equation}
where $\omega$ is the frequency, $k_{h}$ is the horizontal wavenumber ($k_{h}^2
\textnormal{=} k_{x}^2 + k_{y}^2$), $c_{\rm s}$ is the adiabatic sound speed,
$\omega_{\rm ac}$ is the acoustic cut-off frequency, and $N$ is the
Brunt-V\"{a}is\"{a}l\"{a} frequency, explained later in
Equations~(\ref{eq:acutoff}) and (\ref{eq:bruntfreq}).

\begin{figure}
	\centering
	\includegraphics[width=\columnwidth]{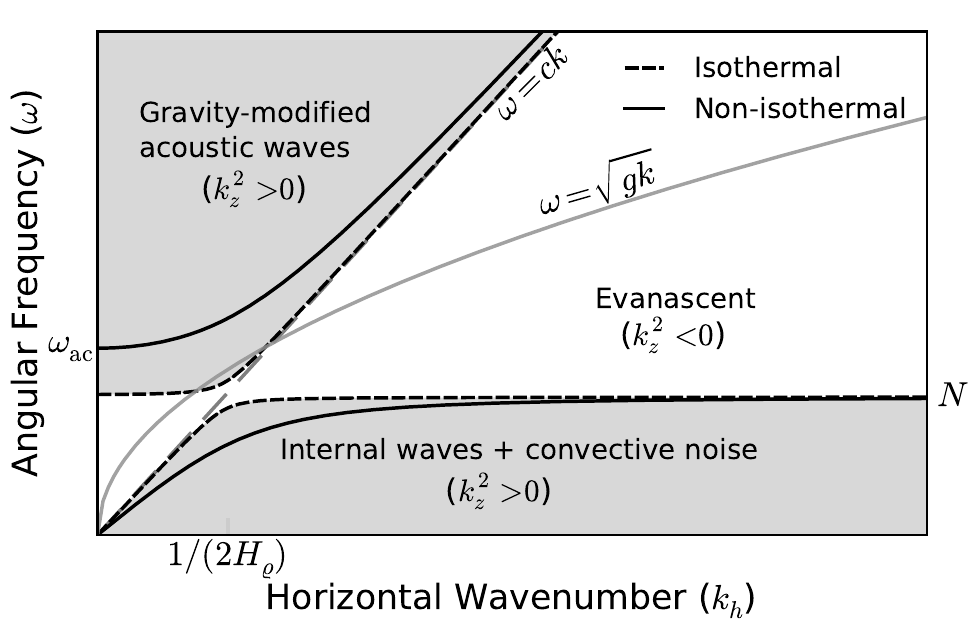}
	\caption{
		Schematic diagram showing different regimes of wave propagation in a
		compressible, gravitationally stratified medium for a given height in
		the atmosphere. The shaded area marks regions of vertical propagation
		of acoustic and the gravity waves. The propagation boundaries separate
		the vertically propagating ($k_{z}^2 \textgreater 0 $) from the
		evanescent ($k_{z}^2 \textless 0 $) solutions. The solid curve
		represents the propagation boundaries obtained from the non-isothermal
		cut-off frequencies defined in Equations~(\ref{eq:acutoff}) and
		(\ref{eq:bruntfreq}). The dashed curves are obtained when we use the
		isothermal approximation for $\omega_{\rm ac}$. The dispersion
		relation for the surface-gravity wave is shown in gray. The long
		dashed gray line corresponds to the dispersion relation for the Lamb
		waves.}
	\label{fig:kwschematic}
\end{figure}

The \textit{local dispersion relation}, given by
Equation~(\ref{eq:disp_relation}), separates the wave-behaviour in the
$k_{h}\textnormal{-}\omega$ diagram also known as the diagnostic diagram. The
two regions of propagation in the $k_{h}\textnormal{-}\omega$ diagram are
obtained by setting $k_{z}^2=0$ in Equation~(\ref{eq:disp_relation}) \cite[see
e.g.,][]{1981NASSP.450..263L}, with the $k_{z}^2\textgreater0$ domain isolating
the vertically propagating solution from the evanescent region
($k_{z}^2\textless0$). A schematic of such a diagnostic diagram for a
compressible, gravitationally stratified medium for a given height in the
atmosphere is shown in Figure~\ref{fig:kwschematic}.

For small $k_{h}$ ($\textless 1/(2 H_{\varrho})$), the lowest frequency with
which a gravity-modified acoustic wave can propagate upward is limited by the
acoustic cut-off frequency 
($\omega_{\rm ac} \textnormal{=} c_{\rm	s}/(2H_{\varrho})$), 
which, for an isothermal atmosphere, is a function of the
sound speed and the density scale height ($H_{\varrho}$), referred to as the
Lamb frequency. However, in the non-isothermal case like that of the solar
atmosphere, the gradients in temperature modify the cut-off frequency. While
there are different expressions for the cut-off frequency, depending on
different representations of the wave equation
\citep{1995A&A...293..586M, 
1998A&A...337..487S}, 
in this paper, we adopt the one due to 
\citet{1984ARA&A..22..593D}, 
viz.,
\begin{equation}
\omega_{\rm ac}^2 = \frac{c_{\rm s}^2}{4 H_{\varrho}^2} \left(1-2\frac{{\rm d}
	H_{\varrho}}{{\rm d} z}\right),
\label{eq:acutoff}
\end{equation}
which is obtained when the wave equation is cast in terms of
$\varrho^{1/2}c_{s}^2 {\nabla\cdot v}$ as the oscillating function. The
difference in the diagnostic diagram between the isothermal and the
non-isothermal case for a particular height in the atmosphere is also shown in
Figure~\ref{fig:kwschematic}.

Internal waves exist below the acoustic cut-off frequency and have horizontal
phase velocity less than the sound speed in the medium. The maximum frequency of
propagation for internal waves is set by the Brunt-V\"{a}is\"{a}l\"{a} frequency
($N$), also called the stratification or buoyancy frequency. For a
non-isothermal atmosphere, it is defined as, \\
\begin{equation}
N^{2} = g\left(\frac{1}{H_{\varrho}} - \frac{1}{\gamma H_{\rm p}}\right),
\label{eq:bruntfreq}
\end{equation}
where, $\gamma$ is the ratio of the specific heats ($c_{P}/c_{V}$). Recalling
that the pressure scale height ($H_{\rm p}$) is equivalent to the density scale
height ($H_{\varrho}$) in an isothermal atmosphere, the expression for the
Brunt-V\"{a}is\"{a}l\"{a} frequency in an isothermal case can be recovered. In
the presence of a magnetic field, the Brunt-V\"{a}is\"{a}l\"{a} frequency can be
further modified, but we do not consider this effect here.

A fluid element vertically displaced from its equilibrium position will
oscillate and emit gravity waves provided the background atmosphere satisfies
the Schwarzschild criterion for stability ($N^2\textgreater 0$). If there are
local departures from the stability criterion due to the overshot material in a
stable stratified surrounding, the fluid element becomes unstable and rises up,
cools down by radiating and falls back, completing the convective cycle. In
observations, the frequency range covering the internal waves is dominated by
the convective noise, but the propagation properties of the internal waves have
been studied by carrying out a phase spectra analysis of these waves.

We have presented the diagnostic diagram and the significance of distinguishing
the two-wave behavior in such a diagram. In the following section, we will look
at the analysis of the simulation data based on this diagnostic diagram.

\section{Spectral Analysis}{\label{s:spectral_analysis}}
The complex cross-spectrum of two real-valued processes: $f(\boldsymbol{x}, t)$,
$g(\boldsymbol{x}, t)$, is defined as:
\begin{mathletters}
	\begin{eqnarray}
	\mathcal{S}_{f,g} (\boldsymbol{k}, \omega) & \equiv & \mathcal{C}_{f,g} (\boldsymbol{k}, \omega) 
	+ i \mathcal{Q}_{f,g} (\boldsymbol{k}, \omega), \nonumber  \\
	& = & \mathcal{F}(\boldsymbol{k}, \omega)~\overline{\mathcal{G}(\boldsymbol{k}, \omega)}.  
	\label{eq:cross_spectra}
	\end{eqnarray}
\end{mathletters}
$\mathcal{F}(\boldsymbol{k}, \omega)$ and $\mathcal{G}(\boldsymbol{k}, \omega)$ are the Fourier
transforms of the two processes, with the overbar representing the complex
conjugate. The real part of $\mathcal{S}$ is known as the co-spectrum
($\mathcal{C}$), and gives the correlation of the in-phase/anti-phase Fourier
components ($\boldsymbol{k}, \omega$) of the two processes. The imaginary part of
$\mathcal{S}$ is known as the quadrature spectrum ($\mathcal{Q}$) and represents
the correlation of the out-of-phase Fourier components between the two processes
\citep{hayashi1982}. These quantities will be further explored in the context of
energy fluxes of the internal waves discussed in
Section~\ref{ss:energy_flux_spectra}.

Using the cross-spectrum, the phase lag or the phase difference between the two
processes is formally given as,
\begin{equation}
\phi_{f,g} (\boldsymbol{k}, \omega) = \tan^{-1} \left[\frac{\mathcal{Q}_{f,g} (\boldsymbol{k},
	\omega)}{\mathcal{C}_{f,g} (\boldsymbol{k}, \omega)}\right],
\label{eq:phase}
\end{equation}
where, $\phi(\boldsymbol{k}, \omega)$ is known as the phase difference spectrum, or
simply the phase spectrum. However, Equation~(\ref{eq:phase}) gives reliable
phases only if the two processes are linearly dependent for a given Fourier
component. The linear dependence of the two processes is measured by the
coherence spectrum ($\mathcal{K}$), defined as,
\begin{equation}
\mathcal{K}_{f,g}^{2}(\boldsymbol{k}, \omega) = \frac{\mathcal{C}_{f,g}^2 (\boldsymbol{k},
	\omega) + \mathcal{Q}_{f,g}^2 (\boldsymbol{k}, \omega)}{\mathcal{S}_{f,f} (\boldsymbol{k},
	\omega)~\mathcal{S}_{g,g} (\boldsymbol{k}, \omega)},
\end{equation}
with $\mathcal{S}_{f,f}$ representing the auto-spectrum of process $f$ and
$\mathcal{S}_{g,g}$ representing the auto-spectrum of process $g$, according to
Equation~(\ref{eq:cross_spectra}). The phase spectra, together with the
coherence spectra give an estimate of the phase-difference between the two
processes, with $\mathcal{K}$=1, when the two processes are linearly related,
and $\mathcal{K}$=0, when no linear dependence exists for the given Fourier
component.

In our analysis, the components of velocity and various other thermodynamic
quantities are extracted from the two models for the entire duration of the
simulation. We then carry out the analysis in the three-dimensional Fourier
space by transforming the data cube of the derived quantities consisting of two
horizontal spatial ($x,y$) and one temporal ($t$) direction, using Fast Fourier
Transform (FFT). This is done so for each horizontal plane of the vertical
coordinate grid (the $z$ axis) to obtain a four-dimensional data set of the
relevant quantities on a ($k_{x},k_{y},\omega,z$) grid. The derived quantities
are then represented on a $k_{h}\textnormal{-}\omega$ diagram for each height
level by azimuthally averaging over the $k_{x}\textnormal{-}k_{y}$ plane. With
the domain spanning 38.4~Mm in the horizontal directions and 8 hours long, we
have a spectral resolution of 0.164~Mm$^{-1}$ in horizontal wavenumber and
138~$\mu$Hz in frequency. The grid resolution of 80~km results in a Nyquist
wavenumber ($k_{\rm Ny} \textnormal{=} \pi/\delta x$) of 39.25 Mm$^{-1}$ of
which we are only interested in horizontal wavenumbers below 8 Mm$^{-1}$, where the bulk of IGWs occur. 
A vertical and horizontal grid constant of respective 20~km and 80~km is sufficient 
	to capture the range of the internal wave spectrum in the models as will be 
	discussed in Sect.~\ref{s:discussion}.
Snapshots from the simulations were taken at 30\,s interval resulting in a
Nyquist frequency ($\nu_{\rm Ny}$) of 16.66\,mHz. Since the
Brunt-V\"{a}is\"{a}l\"{a} frequency in the atmosphere is typically below 5\,mHz,
we show in the following only the analysis up to the frequency range of 8\,mHz.

\subsection{Phase and coherence
spectra}{\label{ss:phase_diff}}
Acoustic waves and internal waves have different polarization properties and
therefore show different behaviour in their phase spectra. Unlike for acoustic
waves, the velocity fluctuations of internal waves and therefore the energy
transport (ray path) of the wave is perpendicular to the wave vector $\boldsymbol{k}$.
Moreover, the wave vector is always directed towards the plane of the source of
perturbation that excited the wave
\citep[see e.g.,][]{sutherland2010}. 
Hence, an internal wave transporting energy
at an angle to the vertical, with an upward component, will have a downward
propagating phase component which shows up as negative phase lag between two
geometrical heights. This behaviour can be clearly identified by computing the
phase spectra obtained from velocity measurements at two different heights. The
diagnostic potential of the phase and coherence diagram was explored in a series
of papers by
\citet{1989A&A...213..423D},
\citet{1989A&A...224..245F},
\citet{1990A&A...228..506D},
\citet{1990A&A...236..509D}, 
and
\citet{1992A&A...266..560D}.
These have been used to separate out the internal wave signature from the low
frequency convective noise.

\begin{figure}
	\begin{center}
		\includegraphics[width=\columnwidth]{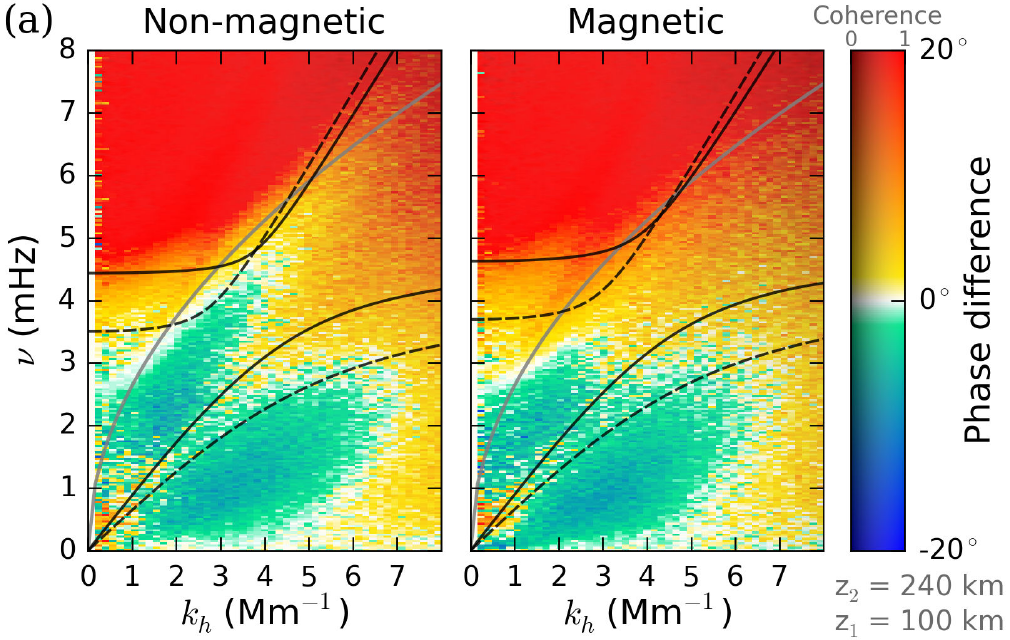}
	\end{center}
	\begin{center}
		\includegraphics[width=\columnwidth]{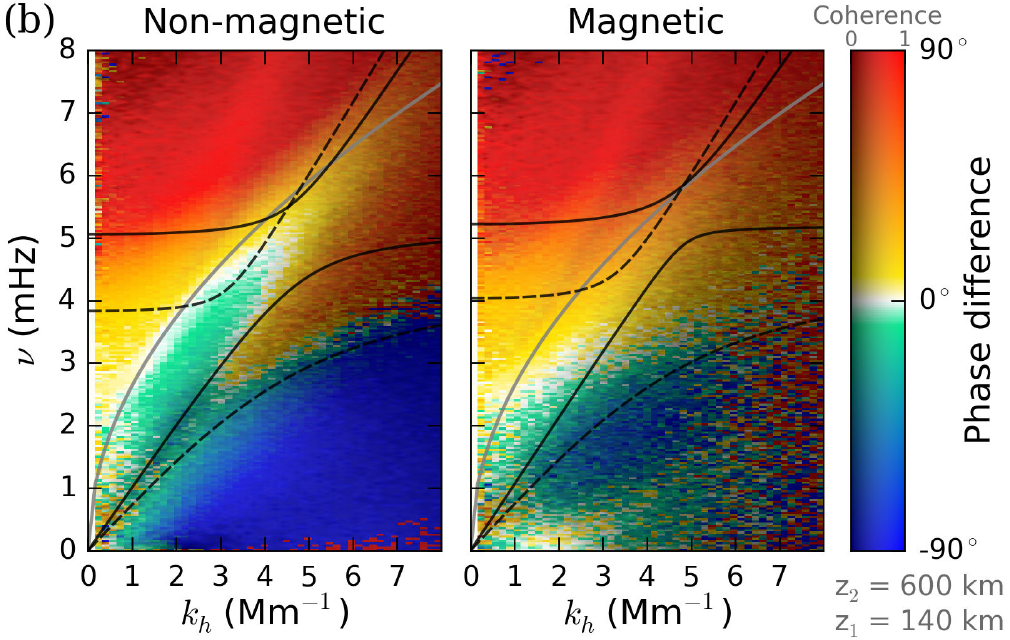}
	\end{center}
	\begin{center}
		\includegraphics[width=\columnwidth]{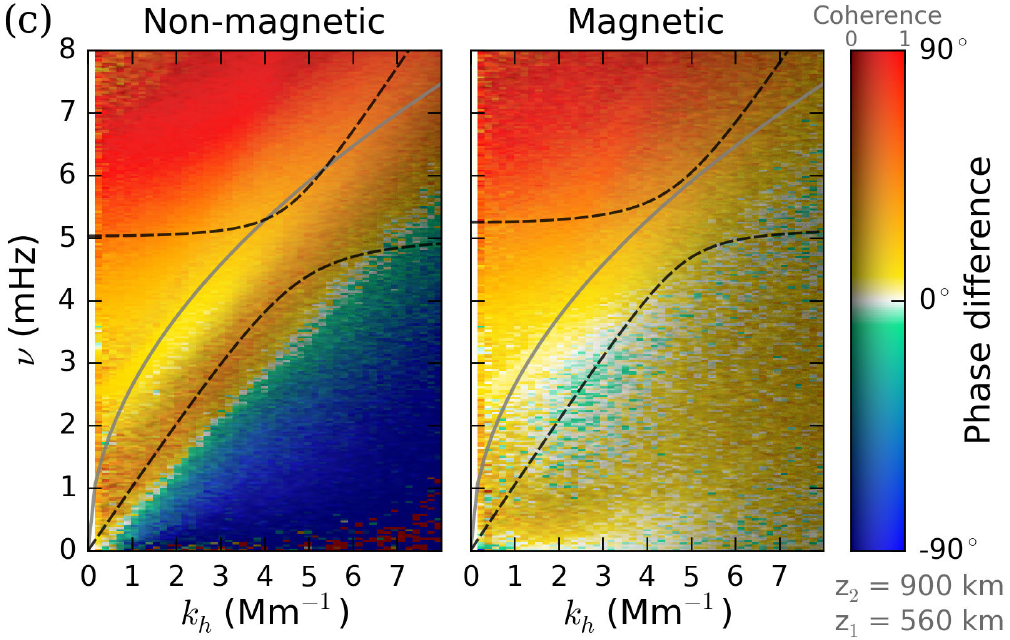}
	\end{center}
	\caption{
		$v_{z}\textnormal{-}v_{z}$ phase spectra estimated between: a)
		$z\textnormal{=}100$\,km and $z\textnormal{=}240$\,km; b)
		$z\textnormal{=}140$\,km and $z\textnormal{=}600$\,km; and c)
		$z\textnormal{=}560$\,km and $z\textnormal{=}900$\,km, for the
		non-magnetic model (left) and the magnetic models (right). The dashed
		black curves represent the propagation boundaries obtained from the
		non-isothermal cut-off frequencies defined in
		Equations~(\ref{eq:acutoff}) and (\ref{eq:bruntfreq}) for the lower
		height and the solid curves correspond to the upper height. The gray
		curve is the dispersion relation of the surface-gravity waves. The
		colors represent the phase differences ($\phi$) and the shading shows
		the coherency ($\mathcal{K}$).}
	\label{fig:phase_diff_3heights}
\end{figure}

In the following, we look at the velocity-velocity ($v$-$v$) phase spectra,
which shows the phase lag between the velocities measured at two different
heights. The $v_{z}$-$v_{z}$ phase spectra are determined from the vertical
component of the velocity for a pair of heights as described in the beginning of
Section~\ref{s:spectral_analysis} and represented in the form of the diagnostic
diagrams. While phase spectra determined from observations of the solar
atmosphere rely on spectral lines formed over a particular height range, in this
work we focus only on phase spectra obtained from pairs of plane parallel,
geometrical height levels. Figure~\ref{fig:phase_diff_3heights} shows the
$v_{z}$-$v_{z}$ phase spectra for pairs of heights for the non-magnetic (left
panels) and for the magnetic (right panels) model of
Table~\ref{tab:model_summary}. In order to better understand the effect of
magnetic fields on the propagation of internal waves, we study the phase spectra
obtained from three carefully selected pairs of heights. These heights are
chosen in such a way that they probe three regions of interest in the magnetic
case. The colors represent the phase differences ($\phi$) and the shading
represents the coherency ($\mathcal{K}$), with corresponding colorbars shown on
the right of the plots. Positive phases (upward) are represented with a
progressively yellow to red color-scale and the negative phases (downward) are
shown with a green to blue color-scale. The shading scale for the coherency is
shown on the top of the colorbar. The gray curve in each plot shows the
dispersion relation of the surface gravity waves. The dashed and solid curves
correspond to the propagating boundaries of the two wave branches at the lower
and the upper height, respectively.

The first pair of heights, $z\textnormal{=}100$\,km and
$z\textnormal{=}240$\,km, lies close to the surface, where the internal waves
are thought to be excited by overshooting convection. In the magnetic model,
this height range probes a gas-dominated part of the atmosphere
($\beta$\textgreater 1, where $\beta$ is the ratio of the gas pressure to the
magnetic pressure) . The diagnostic diagram of these two heights is shown in
Figure~\ref{fig:phase_diff_3heights}a, where we see that both models have
generated significant amounts of internal waves, which show up as downward
phases in the internal gravity wave-regime of the diagnostic diagram (the green
area below the lower dashed curve that show phase difference of around
$-10^{\circ}$ over a height difference of 140\,km). Although, the generation of
the internal waves and how magnetic fields influence the generation is of great interest,
we defer such a study to a later paper. Here, we focus only on the propagation
properties of these waves in the presence of magnetic fields. As can be seen,
the downward phases are restricted to the region below the dashed curve,
suggesting that the excited internal waves are propagating only below the
boundary determined by the lowest Brunt-V\"{a}is\"{a}l\"{a}
frequency (in this case, the $N$ of the lower
height). 

The two spectra of the excited internal waves in
Figure~\ref{fig:phase_diff_3heights}a are qualitatively the same regardless of
whether being generated in the convective or magneto-convective model. It should
be noted that the magnetic model, however, inhibits surface gravity waves, the
spectrum of which is clearly seen as a green ridge extending along the gray
curve in the non-magnetic model. This could be due to the fact that the magnetic
fields in the simulation box are predominantly vertical so that the propagation
of the nearly horizontal surface gravity waves are hindered by their presence.

Now we turn to Figure~\ref{fig:phase_diff_3heights}b, the second pair of heights
($z\textnormal{=}140$\,km and 600\,km), which are still within predominantly gas
dominated regions ($\beta\textgreater 1$). But, in the atmosphere that these
heights probe, the surfaces of constant plasma-$\beta$ are rugged with
occasional strong magnetic fields dipping the plasma-$\beta$ surfaces. The
non-magnetic model shows the signature of internal waves with the downward
phases with phase differences of around $-90^{\circ}$ over a height difference
of 460\,km. In the magnetic model they are significantly reduced, suggesting
that the magnetic fields have a major influence on the internal waves as they
propagate upwards. Here again, the negative phase difference, and therefore the
propagating region in the diagnostic diagram is mainly below the boundary set by
the $N$ of the lower height. Also note that the coherence has reduced as evident
from the increased shading for the larger wavenumbers, since we are probing
heights separated by a larger distance. The surface-gravity waves (ridge along
the gray curve), on the other hand, are still present in the non-magnetic model,
but they are completely absent in the magnetic model.

Figure~\ref{fig:phase_diff_3heights}c refers to the third pair of heights
($z\textnormal{=}560$\,km and 900\,km), where the first height is in a gas
dominated region ($\beta\textgreater 1$) and the second height is in the
magnetic field dominated region ($\beta\textless 1$). We see that most of the
internal waves are absent in the magnetic model (phase difference of
$0^{\circ}$ over a height difference of 340\,km). Some regions of the diagnostic
diagram in the internal wave regime of the magnetic model also show positive phase differences (upward
propagating phases) of around $10^{\circ}$. According to their polarisation
properties, this suggests that the wave energy is propagating downwards in the
atmosphere.

In summary, the non-magnetic case shows a strong negative phase difference in
the internal wave region in all three pairs of heights, while the magnetic case
shows a clear signature of upward propagating internal waves for the pair of
heights in the lower atmosphere and mostly zero to positive phase differences in
the upper atmosphere. From the above analysis, we observe that the presence of
nearly vertical magnetic fields influences internal waves and it results in
their suppression or partial reflection in the atmosphere. There are several
ways by which internal waves can behave this way, and we explore some of these
factors in Section~\ref{s:discussion} to understand the behaviour that we see in
our simulation.

\subsection{Energy flux spectra}{\label{ss:energy_flux_spectra}}
The phase spectrum analyses show that in the case of the magnetic model the
internal waves are absent or even show a positive phase difference because they
propagate down in the higher layers. This means that in this case they are
either destroyed or reflected back and are transporting their energy downwards,
unlike the acoustic waves which mainly transport their energy upwards in the
atmosphere. An estimate of the energy flux spectra can shed some light on the
actual energy transport by internal waves in the presence of magnetic fields.

A propagating wave transports energy to the far field, when pressure and
velocity oscillate in-phase. In order to estimate the vertical component of the
linearized mechanical energy flux of these waves, we look at the co-spectrum of
the pressure fluctuations, $\Delta p$, and the vertical component of the
velocity, $v_{z}$
\citep{2001wafl.book.....L}, 
averaged over one wavelength (1/2 factor). As described at the beginning  of
Sect.~\ref{s:spectral_analysis}, the co-spectrum gives us the in-phase
cross-spectrum, which in this case, is the active mechanical energy flux
transported by the waves,
\begin{mathletters}
	\begin{eqnarray}
	F_M (\boldsymbol{k}, \omega)
	& = & \frac{1}{2} \mathcal{C}_{\Delta p, v} (\boldsymbol{k}, \omega), \nonumber \\
	& = & \frac{1}{2} {\rm Re} [\Delta p(\boldsymbol{k}, \omega)\,\overline{v(\boldsymbol{k}, \omega)}]. 
	\label{eq:energy_flux}
	\end{eqnarray}
\end{mathletters}

\begin{figure}
	\begin{center}
		\includegraphics[width=\columnwidth]{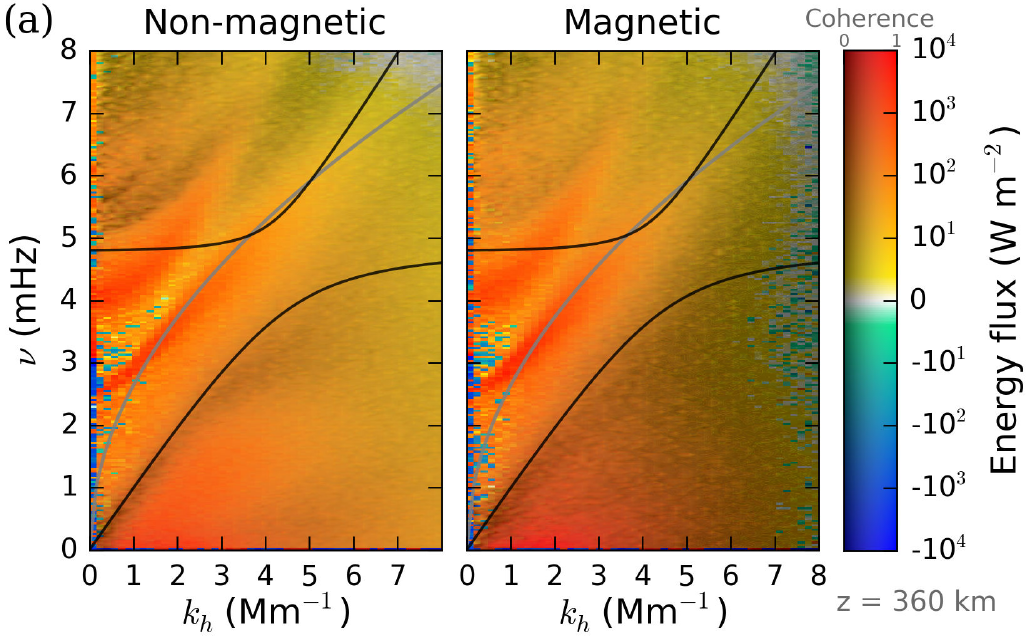}
	\end{center}
	\begin{center}
		\includegraphics[width=\columnwidth]{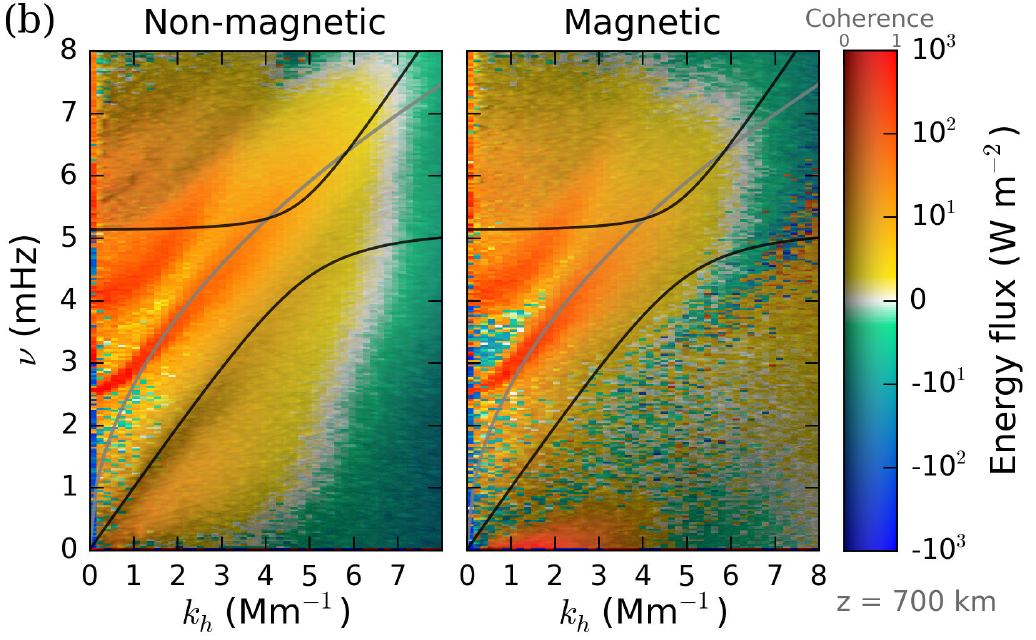}
	\end{center}
	\caption{
		Energy flux spectra at heights of a) $z\textnormal{=}360$\,km and b)
		700\,km of the non-magnetic model (left) and the magnetic model
		(right). The solid black curves represent the propagation boundaries
		obtained from the non-isothermal cut-off frequencies defined in
		Equations~(\ref{eq:acutoff}) and (\ref{eq:bruntfreq}). The gray curve
		is the dispersion relation of the surface-gravity waves.}
	\label{fig:energy_2heights}
\end{figure}

The energy flux, $F_{M}$, calculated using Equation~(\ref{eq:energy_flux}) in
the $k_{x}$-$k_{y}$ plane is then azimuthally averaged and represented on the
diagnostic diagram. Figure~\ref{fig:energy_2heights}a and
\ref{fig:energy_2heights}b shows the energy flux spectra computed at a height of
$z\textnormal{=}360$\,km and $z\textnormal{=}700$\,km, respectively. Positive
values correspond to upward flux and negative values correspond to downward
flux. The energy flux spectra computed for $z\textnormal{=}360$\,km (see
Figure~\ref{fig:energy_2heights}a) show that both the acoustic and the internal
waves transport their energy upwards in both the magnetic and non-magnetic
model. When we look at the energy flux spectra at $z\textnormal{=}700$\,km (see
Figure~\ref{fig:energy_2heights}b), it is clear that the internal waves in the
non-magnetic model still carry a positive flux, which means they are propagating
and transporting energy predominantly upwards. However, the magnetic model shows
a mixture of positive and negative energy flux in the gravity wave regime (in
locations where there is a negative phase difference in the right panel of
Figure~\ref{fig:phase_diff_3heights}b), suggesting that at this height, the
waves are propagating in both directions up and down, and thus energy is
transported in both directions. The upward propagating waves are probably the
one that are generated in the lower atmosphere, and the downward propagating
waves are the one reflected from the top layer of the atmosphere.

In this  work, we have not attempted to compute the Poynting flux from the
magnetic model, as we cannot do a comparative study with the non-magnetic model.
Future work will explore the emergent Poynting flux by comparing different
magnetic models.

\section{Discussion}{\label{s:discussion}}
We now focus our attention on explaining the behaviour of internal waves that
are seen in the numerical models, particularly the absence or the downward
propagation in the magnetic model, which is also partially evident from the
energy flux spectra. We explore different factors that may affect the
propagation of internal waves in a realistic atmosphere. All the factors
considered below can restrict the possible height range over which internal
waves can occur in the solar and, generally, in stellar atmospheres. We start by
looking at the differences in the height dependence of the diagnostic diagram in
both models and how this affects the propagation of internal waves, followed by
the influence of radiative damping and non-linear effects and finally the
presence of magnetic fields. We will see that, while the lower and upper
limiting boundaries of the internal wave cavity are determined by the radiative
damping effects and flow parameters, respectively, the propagation within the
allowed domain is strongly influenced by magnetic fields.

We note that the effect of numerical diffusion becomes important at the level of a 
	couple of grid cells only. The artificial diffusion in {CO$^{\rm 5}$BOLD} is invoked at 
	shock fronts or for waves with large amplitudes, where strong gradients of 
	velocity exist. Since gravity waves do not shock or do not steepen very much, 
	they are not affected by artificial numerical diffusion; it influences waves of short 
	wavelengths only, which, however, are irrelevant in this study since we see the 
	effects of magnetic fields mainly at long wavelengths. Also, current observations 
	of IGWs do not have the spatial resolution to detect power at such short wavelengths. 
	On the other hand, since in our models the horizontal wave number of the propagating IGWs 
	is smaller than 7 Mm$^{-1}$ (see Fig.~\ref{fig:phase_diff_3heights}a), which corresponds to wavelengths larger than 
	$\approx 1000$\,km, they are well resolved with the horizontal grid spacing of 80\,km. 
	Likewise, in the vertical direction, Fig.~\ref{fig:phase_diff_3heights}a together with Eq.~(\ref{eq:disp_relation}) tells us that
	$k_z < 40$\,Mm$^{-1}$ corresponding to wavelengths larger than $\approx 160$\,km,
	which are well resolved with the present vertical grid spacing of 20\,km.

\subsection{Variation of the diagnostic diagram with height}{\label{ss:heightvariation}}
In the case of a convectively stable, uniformly stratified atmosphere, $N^2$ is
positive and constant and an internal wave can freely propagate throughout the
atmosphere. However, in a more realistic atmosphere like the one we simulate,
$N$ varies with height. Variations or discontinuities in $N$ result in partial
reflection or trapping (ducting) of internal waves within the domain. As we have
seen in Section~\ref{ss:kwdiagram}, a spectral band of evanescent disturbances
(white region in Figure~\ref{fig:kwschematic}) separates the gravity-modified
acoustic waves from the internal gravity waves (gray region in
Figure~\ref{fig:kwschematic}). Waves with a specific ($k_{h}$, $\omega$) that
fall in either of these two gray regions in the diagnostic diagram, of a certain
height, have oscillatory solutions at that particular height and propagate as
waves with their characteristic nature. All other combinations of ($k_{h}$,
$\omega$) are evanescent in the atmosphere. The parameters that set these limits
are mainly $\omega_{\rm ac}$ and $N$, which vary as a function of height in the
real solar atmosphere leading to changing wave behaviour, i.e, a changing
diagnostic diagram with height.

\begin{figure}
	\centering
	\includegraphics[width=\columnwidth]{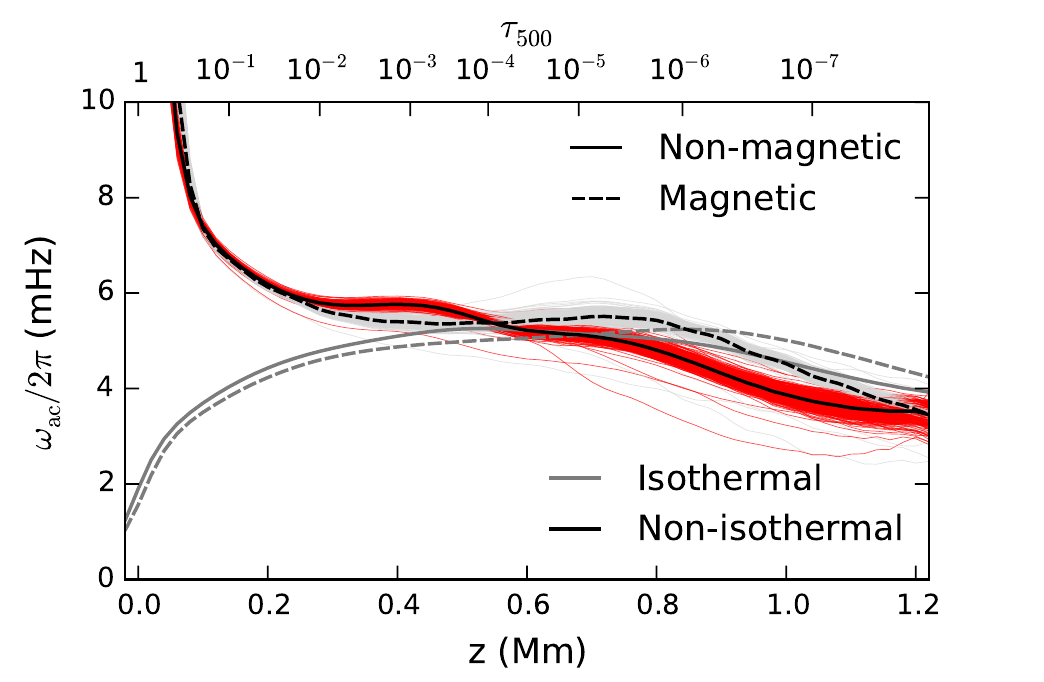}
	\caption{
		Temporally and horizontally averaged isothermal (gray curves) and the non-isothermal (black curves)
		acoustic cutoff ($\omega_{\rm ac}$) frequency as a function of
		height in the non-magnetic (dashed) and the magnetic (solid) model
		above $z\textnormal{=}0$\,km. The gray and red scatter indicate the
		temporal variation of the non-isothermal acoustic cut-off for the
		non-magnetic and the magnetic simulations, respectively.}
	\label{fig:acutoffs}
\end{figure}

\begin{figure}[ht]
	\centering
	\includegraphics[width=\columnwidth]{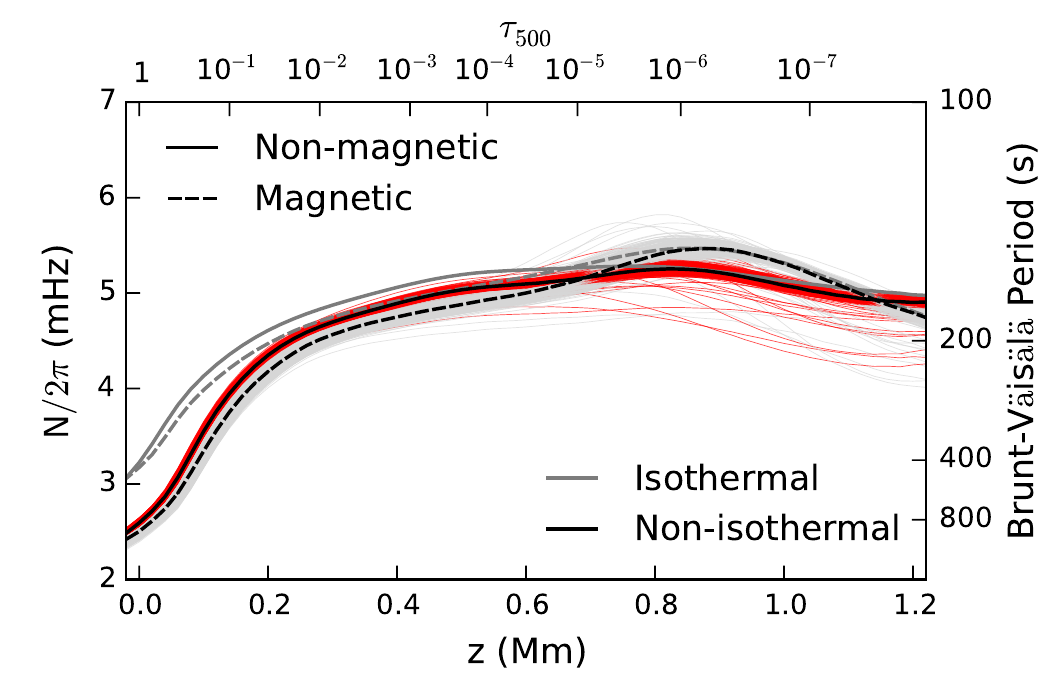}
	\caption{
		Temporally and horizontally averaged isothermal (gray curves) and the
		non-isothermal (black curves) Brunt-V\"{a}is\"{a}l\"{a} frequency as a
		function of height in the non-magnetic (dashed) and magnetic (solid)
		models above $z\textnormal{=}0$\,km. The gray and red scatter indicate
		the temporal variation of the non-isothermal Brunt-V\"{a}is\"{a}l\"{a}
		frequency for the non-magnetic and the magnetic simulations,
		respectively.}
	\label{fig:nbrunts}
\end{figure}

Figure~\ref{fig:acutoffs} shows the time-averaged $\omega_{\rm ac}$ as a
function of height in the two simulations that are presented in this paper
(black curves). The variation of $\omega_{\rm ac}$ is the result of the changing
temperature and stratification. The $\omega_{\rm ac}$ for the iso-thermal case
is shown in gray which takes into account only the local sound speed and density
scale height.
Figure~\ref{fig:nbrunts} shows the time-averaged $N$ as a function of height in
the two simulations. The time-averaged $N$ for the isothermal case is shown in
gray. In both figures, the gray and red scatter show the temporal variation of
the non-isothermal value for the non-magnetic and magnetic models, respectively.

In order to fully understand the propagation and transport of energy by the two
types of waves, it is important to know the local diagnostic diagram as a
function of height and thus the critical frequencies as a function of height, as
shown in Figures~\ref{fig:acutoffs} and \ref{fig:nbrunts}. Oscillating solutions
to the wave equation for a particular ($k_{h}$, $\omega$) may exist over the
entire domain or only for a particular range of heights. A wave at a particular
height with a frequency that falls in the white region in
Figure~\ref{fig:kwschematic} is partially \textit{reflected} at the respective
limits, as discussed in connection with Equation~(\ref{eq:disp_relation}),
beyond which it becomes \textit{evanescent}. If such a limit exists at another
height for the same wave, and a propagating wave solution exist for the region
between these two heights, then the wave is said to be \textit{trapped}. On the
other hand, if oscillatory solutions exists on either side, then the waves can
\textit{tunnel} through this barrier. Following the above criterion for the
range of wavelengths present in our simulation, the diagnostic diagram can be
separated into different regions for each branch of the acoustic-gravity
spectrum.

According to Figure~\ref{fig:nbrunts} it is clear that the propagating branch of
internal waves occupy nearly the same region of the $k_{h}\textnormal{-}\omega$
diagram in both models, because $N$ as a function of height is almost identical.
A wave that propagates into a region where it has no oscillatory solution is
partially reflected back towards the propagating region, the rest becoming
evanescent on the opposite side. In our models, these reflecting surfaces for
the internal waves occur in the low photosphere where $N$ sharply drops with
depth\footnote{
\citet{1981ApJ...249..349M} 
considered a 1D atmosphere with effects of ionization and external forcing due
to ``turbulent pressure'' which causes a decrease in $N$ with height having the
consequence that the bottom of the chromosphere acts as a reflecting layer for
waves propagating upwards.} (see Figure~\ref{fig:nbrunts}). Trapped internal
waves in our model occupy a very small region in the $k_{h}\textnormal{-}\omega$
diagram with frequencies close to the maximum $N$ in the entire box. Since these
waves have frequencies close to $N$, their phases propagate almost horizontally,
transporting their energy upwards, which makes them important for energy
transport to the upper atmosphere. However, the range of frequencies that are
trapped is very small in both our models, lying within the concave stretch of
$N$ from $z\textnormal{=}$0.4\,Mm to 1.2\,Mm in Figure~\ref{fig:nbrunts}. This
is small compared to previous work, which considered a larger height range with
a sharp decrease in $N$ with height.

From Figure~\ref{fig:nbrunts} it is evident that the reflection that we observe
in the magnetic model cannot be not due to the variation of $N$ with height
because $N$ remains nearly constant higher up in the atmosphere. In our specific
case, we have only the lower part of the atmosphere acting as a reflecting layer
for the internal gravity waves propagating downwards and the non-magnetic and
magnetic model show a similar variation of $N$ with height.

\subsection{Radiative damping}{\label{ss:damping}}

Internal waves are thought to be generated by overshooting convection into the
stably stratified layer above. While, the
lower boundary for the waves to exist is determined by the positivity of $N^2$
(which is the condition for a convectively stable region), radiative effects
play an important role in damping the waves higher up in the atmosphere. Near
the surface, the radiative relaxation time, $\tau_{\rm{rad}}$, defined in the optically
thin limit as
\citep{1957ApJ...126..202S},
\begin{equation}
\tau_{\rm{rad}} = \frac{\varrho c_{V}}{16\kappa\sigma T^3},
\end{equation}
drops sharply to values of seconds. Thus temperature fluctuations are smoothed
out on comparable timescales. However, $\tau_{\rm{rad}}$ rapidly increases with height
again, so that radiative effects have no influence on the propagation of
internal waves in the layers above the mid-photosphere. Internal waves with
periods larger than $\tau_{\rm{rad}}$ are destined to be strongly damped in the near
surface layers. The effect of radiative damping on internal waves has been
extensively studied by
\citet{1982ApJ...263..386M}, 
who consider a simple linear height dependent Newtonian cooling and assume
different initial energy fluxes for the waves.

\begin{figure}[ht] 
	\centering 
	\includegraphics[width=\columnwidth]{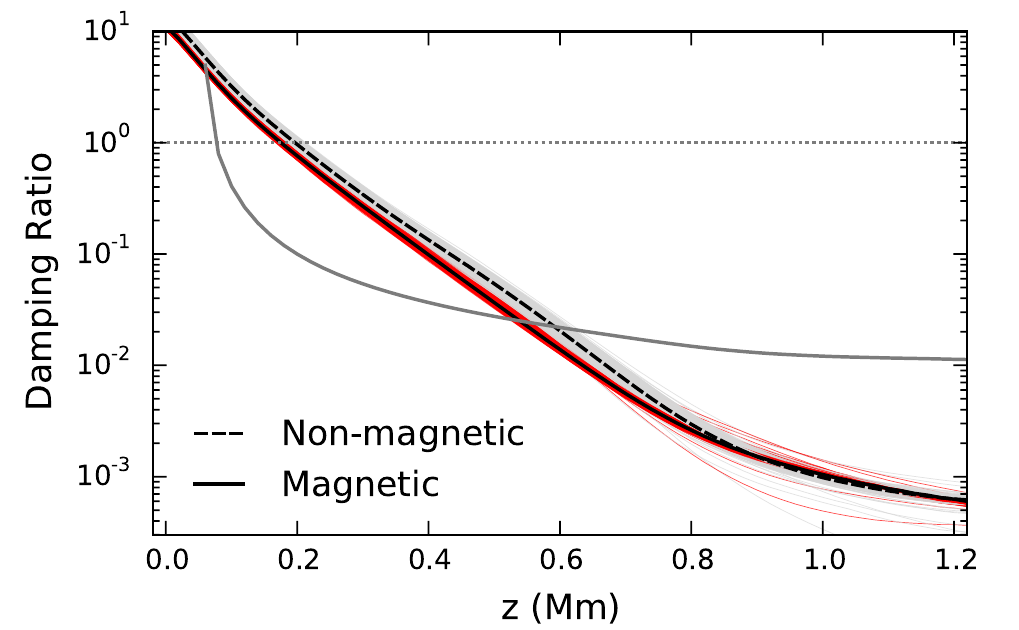}
	\caption{
		Damping ratio as a function of height in the non-magnetic (dashed)
		and magnetic (solid) model. The gray and red scatter indicate the
		temporal variation of the damping ratio for the non-magnetic and the
		magnetic simulation, respectively. The gray curve represents the
		approximation used by 
		\citet{1982ApJ...263..386M}.}
	\label{fig:radiative}
\end{figure}

The damping ratio, $1/2N\gamma\tau_{\rm{rad}}$, characterises the effect of radiative
damping of internal waves. Figure~\ref{fig:radiative} shows the damping ratio as
a function of height in both our models. Also shown in gray is the approximation
used by
\citet{1982ApJ...263..386M}
for comparison. It can be clearly seen from the plot that the gravity waves
undergo heavy radiative damping below a height of 0.2\,Mm, where the damping
ratio is above 1. However, the waves are unaffected by radiative damping higher
up in the atmosphere.

In the lower atmosphere, it is clear from the phase spectra that we still see
signatures of upward propagating internal waves, despite strong radiative
damping. It seems that the internal wave flux generated by the convective
overshooting is still strong enough so that a significant amount of internal
waves survive (see Figure~\ref{fig:phase_diff_3heights}a) in regions where the
damping ratio is above 1. Non-local radiative transfer can have an inverse
effect in the sense that instead of smoothing, the spatial temperature
fluctuations are enhanced, as was conjectured by
\citet{1982ApJ...263..386M}
which needs to be further investigated.

\subsection{Non-linear interaction}{\label{ss:nonlinear}}
Internal waves dissipate their energy by breaking into turbulence. In a large
eddy simulation like the one that we carry out here, wave breaking is very
limited. Nevertheless, it is worthwhile to have an estimate of the effect of
different processes that may lead to the breaking of internal waves into
turbulence, or forming critical layers. A `critical level' is defined as the level at which 
	the mean flow speed becomes comparable to the horizontal phase speed of the 
	wave. The most important among them is the effect of a background flow, like the presence of a strong shear flow or
vorticity. In the case of a background plane-parallel shear flow, the height at
which the horizontal phase speed becomes comparable to the background flow
speed, will act as a critical layer resulting in the reflection of waves. The
importance of shear flows for gravity waves can be characterized by the
Richardson number (${\rm Ri}$), defined as,
\begin{equation}
{\rm Ri} = {N^2}/{\left(\frac{{\rm d} v_{h}}{{\rm d} z}\right)^2},
\end{equation}
where $v_h$ is the horizontal component of the velocity. The estimated value of
${\rm Ri}$ in our model atmosphere is everywhere larger than 0.25
\citep[see e.g.,][]{1988PApGe.126..103L}, 
suggesting that the atmosphere is dynamically stable and shear flows that are
strong enough to lead to dynamical instabilities do not exist.

\begin{figure}[ht] 
	\centering 
	\includegraphics[width=\columnwidth]{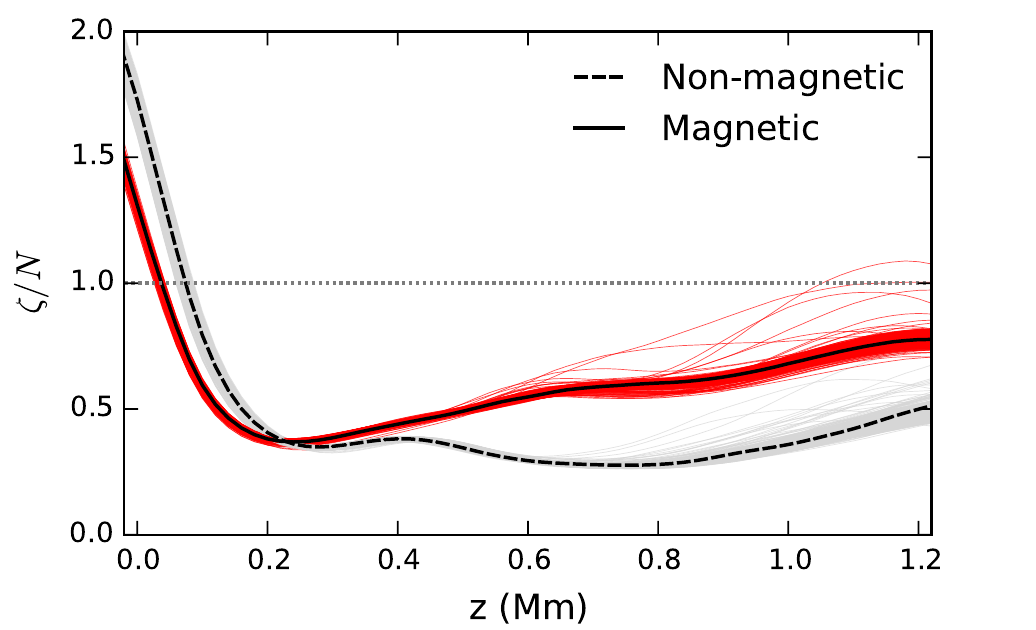}
	\caption{
		Non-linearity parameter ($\zeta/N$) as a function of height in the
		non-magnetic (dashed) and magnetic (solid) models. The gray and red
		scatter indicate the temporal variation of the ratio $\zeta/N$ for the
		non-magnetic and the magnetic simulation, respectively.}
	\label{fig:non-linear}
\end{figure}

Another stability condition considered by 
\citet{1981ApJ...249..349M} 
is the ratio of the wave vorticity, $\zeta$, and $N$.
Figure~\ref{fig:non-linear} shows the ratio of the average fluid vorticity and
$N$ as a function of height. We find that the ratio, $\zeta/N$, is small in both
models above 0.1\,Mm, suggesting that instabilities do not develop as a result
of the flow vorticity in our models. Note however, that $\zeta/N$ is larger and
increases with height in the magnetic model compared to the non-magnetic model,
probably because of the generation of vorticity by the magnetic field in the
low-$\beta$ regime
\citep{2011A&A...526A...5S,
2012ASPC..456....3S,
2012Natur.486..505W}.
We also observe that the vortices in the non-magnetic model near the surface are
larger compared to the magnetic model as also reported in observations by
\citet{2016ApJ...824..120S}.

We recall that the simulations presented in this paper were carried out on a
coarse grid of 80\,km cell size in the horizontal directions which only
marginally captures the development of strong vortical flows. Having a higher
spatial resolution may likely result in vortical flows having a stronger effect
on the internal waves. In fact, high-resolution simulations with a smaller box
size than the one presented here shows that $\zeta/N$ rises above 1 in the top
layers where magnetic fields are present. As can be seen in the
Figure~\ref{fig:non-linear}, there are instances when the $\zeta/N$ increases
and strides above 1 close to the top boundary in the magnetic model. This
implies that vortical motions must be considered a possible reason why internal
waves are absent in the magnetic model.

\subsection{Linear mode coupling}{\label{ss:coupling}}

The presence of magnetic fields itself may play a significant role in modifying
the nature of internal waves in places where they exist.
\citet{2010MNRAS.402..386N,
2011MNRAS.417.1162N} 
considered internal wave propagation in a VAL-C solar reference atmosphere,
containing a uniform magnetic field with different field inclinations. Using
generalized ray theory and with the help of linear simulations, they show that
the internal waves are reflected within the region where plasma
$\beta$\textgreater1, and convert to downwardly propagating slow waves
(predominantly magnetic in nature). The presence of strongly inclined fields
(with an inclination of 80$^\circ$ or more) in these regions can modify the
waves and convert them to acoustic (in case of 2D) or Alfv\'{e}n waves (in case
of 3D) and guide them along the field lines with radiative damping playing only
a minor role
\citep{2011MNRAS.417.1162N}.

\begin{figure}[ht] 
	\centering 
	\includegraphics[width=\columnwidth]{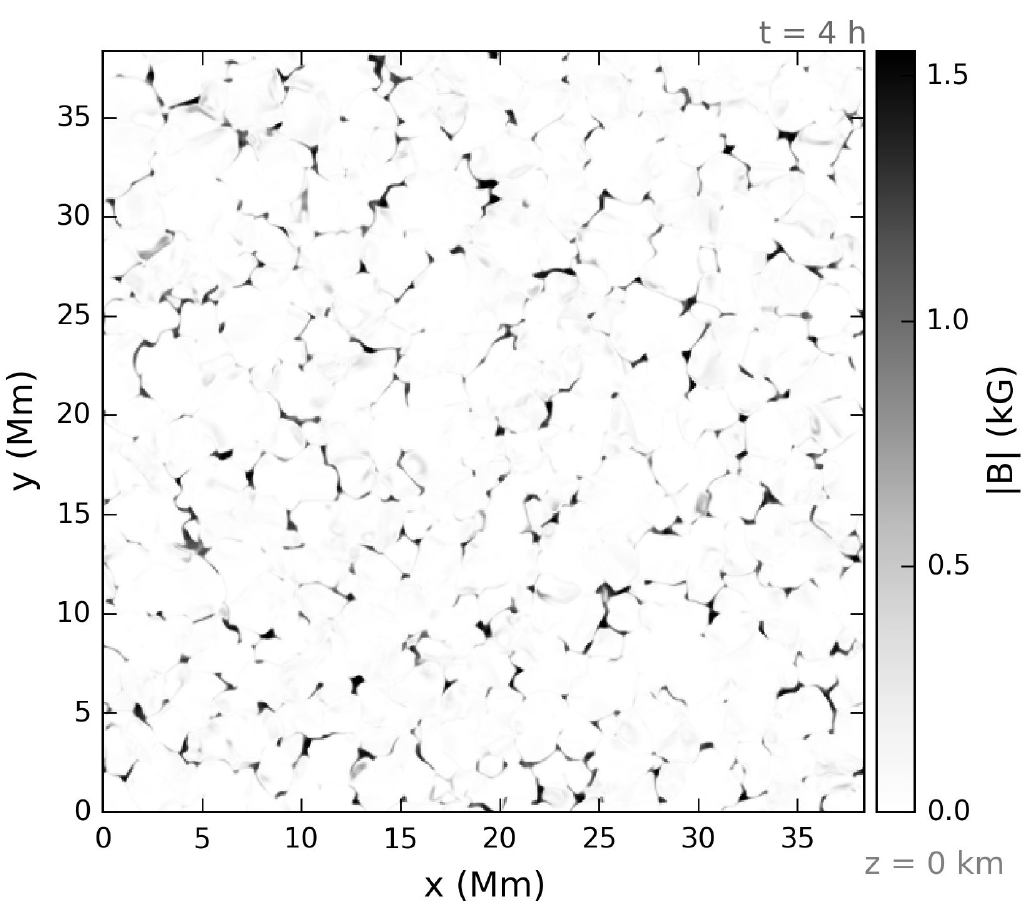}\\
	\caption{
		Snapshot of the absolute magnetic field strength, $|B|$, at
		$t\textnormal{=}4$\,h in the simulation, taken at
		$z\textnormal{=}0$\,km.}
	\label{fig:mag_field_snapshot}
\end{figure}

\begin{figure}[ht] 
	\centering 
	\includegraphics[width=\columnwidth]{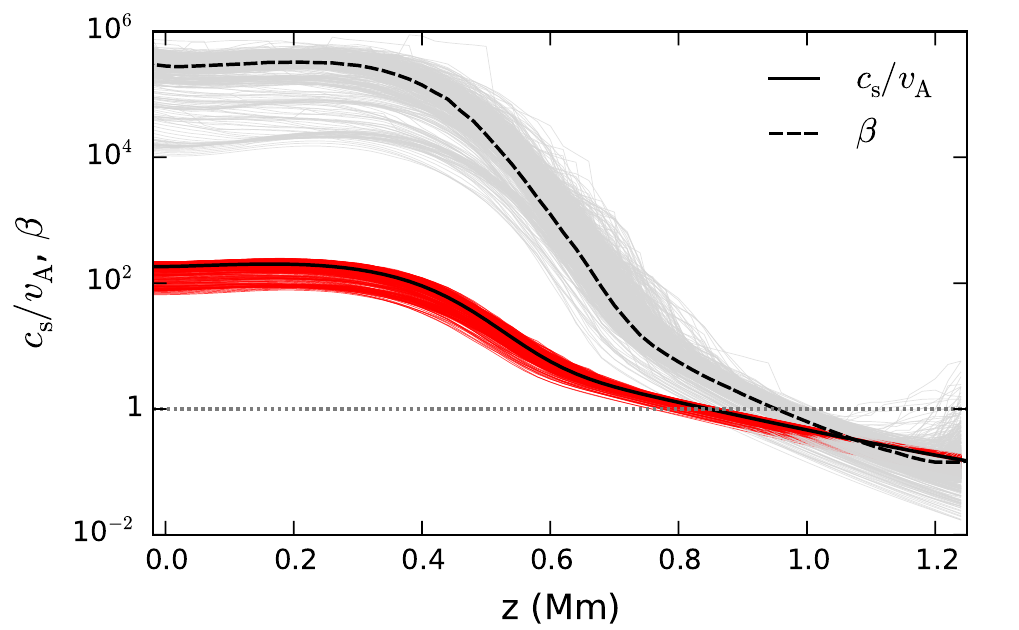}\\
	\caption{
		Temporally and horizontally averaged ratio $c_{\rm s}/v_{\rm A}$ (solid curve)
		and plasma $\beta$ (dashed curve) in the magnetic simulation. The red and gray scatter show the temporal variation of
		$c_{\rm s}/v_{\rm A}$ and plasma $\beta$ , respectively.}
	\label{fig:beta_csva}
\end{figure}

In more realistic simulations like the one we consider in this paper, it is
difficult to specify an average height of the plasma $\beta\textnormal{=}1$
surface or a characteristic inclination of the magnetic fields. The magnetic
fields are continuously shuffled and reformed in the inter-granular lanes
forming a complex structure as shown in Figure~\ref{fig:mag_field_snapshot}. In
order to show how the plasma $\beta\textnormal{=}1$ surface or the magnetic
field inclination vary, we compute the average values of $\beta$, the sound
speed, $c_{\rm s}$, the magnitude of the Alv\'{e}n velocity, $v_{\rm A}$, the
vertical component, $B_{\rm v}$, and the horizontal component of magnetic field,
$B_{\rm h}$, given as $B_{\rm h}^2\textnormal{=}B_{\rm x}^2+B_{\rm y}^2$ over
the entire simulation run as a function of height. Figure~\ref{fig:beta_csva}
shows the plasma $\beta$ (dashed) and the ratio of $c_{\rm s}$ to $v_{\rm A}$
(solid), as a function of height, averaged over horizontal planes and in time
over the entire simulation. Also shown is the temporal scatter of 
$c_{\rm	s}$/$v_{\rm A}$ (light gray) and of plasma $\beta$ (red). From
Figure~\ref{fig:beta_csva}, it is evident that the domain below
$z\textnormal{=}0.8$\,Mm is gas dominated, although there are localized regions
of strong magnetic field that dip the $\beta$ surface down to $z\textless 0$.
According to
\citet{2010MNRAS.402..386N}, 
internal waves in our model are less likely to be present above
$z\textnormal{=}0.7$\,Mm as most of them will undergo conversion to slow
(predominantly magnetic) waves and reflect back before reaching this height.

\begin{figure}[ht] 
	\centering 
	\includegraphics[width=\columnwidth]{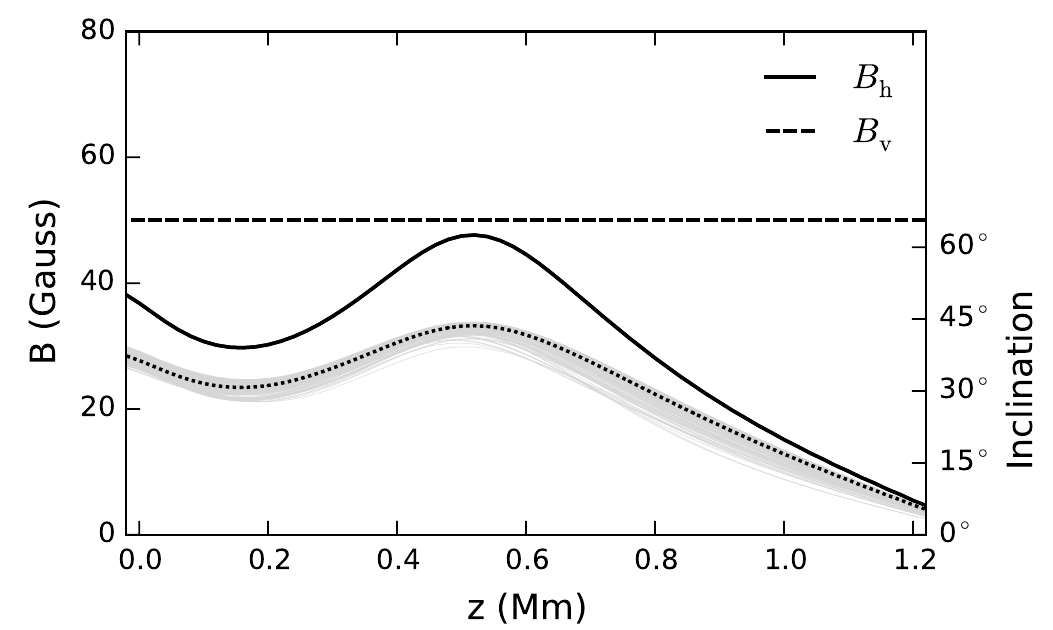}\\
	\caption{
		Temporally and horizontally averaged components of the magnetic
		field in the magnetic simulation. The vertical field is shown as a
		dashed curve and the horizontal field is shown as a solid curve. The
		dotted curve shows the average inclination of the field from the
		vertical, and the gray curves show the temporal scatter of the average
		inclination.}
	\label{fig:b_incl}
\end{figure}

Our simulation also shows a significant horizontal component of the magnetic
field at photospheric heights, in agreement with recent observations of the
solar atmosphere
\citep{2008ApJ...672.1237L,
2012ApJ...751....2O}. 
Figure~\ref{fig:b_incl} shows the average horizontal (solid curve) and vertical
component (dashed) of the magnetic field along with the average field
inclination (dotted) and its temporal scatter shown in gray. The vertical
component of the magnetic field dominates in the entire domain mainly due to the
relatively strong (50\,G) uniform vertical field, of the initial configuration.
However, the fields tend to be inclined around 0.5\,Mm, with a maximum average
inclination of $40^{\circ}$, which can act as a portal for internal
waves to escape into the layers above and convert to acoustic and Alfv\'{e}n
waves. This conversion is highly dependent on the field angle and the model that
we have does not have strongly inclined fields (fields above an inclination of
80$^\circ$) to facilitate this pathway. From the phase spectrum analysis, we do
not see a strong transmission of the internal waves into the upper atmosphere
(see Figure~\ref{fig:phase_diff_3heights}b,c right panels). We can conclude that
most of the waves sense the $c_{\rm S}\textnormal{=}v_{\rm A}$ surface and are
reflected back within the high-$\beta$ region, but we cannot say if it is due to
mode coupling or non-linear shear flow interaction.


\section{Summary and Conclusion}{\label{s:conclusion}}
Internal gravity waves in the solar atmosphere are thought to be generated
mainly by the overshooting of convective matter into the stably stratified
atmosphere lying above. Strong radiative cooling in the immediate vicinity of
the solar surface causes these waves to quickly damp, but they are believed to
be present higher up in the atmosphere where the radiative timescales are large.
Theoretical studies show that the flow field higher up in the atmosphere may
lead to the breaking of internal waves to turbulence resulting in a complete
dissipation of their energy in the mid-chromosphere, before even reaching to
coronal heights. Additional complications are brought about by the presence of
magnetic fields in this region, questioning their ability to transport energy in
the solar atmosphere at all. A clear understanding of the gravity-wave phenomena
occurring in the lower solar atmosphere requires a comprehensive treatment in
three dimensions, including the effects of magnetic fields, non-local radiative
transfer and realistic equation of state.

In this paper, we have presented a study of the acoustic-gravity wave spectrum
emerging from a realistic simulation of solar convection. A purely hydrodynamic
and a MHD simulation were carried out to highlight the effect of the magnetic
fields on the propagation of internal waves. The generated internal waves in
both models are studied in the spectral-domain by looking at the emergent phase
spectra between two heights in the atmosphere and estimating the energy flux
spectra. These studies were carried out in the light of the observations by
\citet{2008ApJ...681L.125S} 
that the gravity waves are suppressed at locations of magnetic flux. These
authors assumed that the suppression is a result of mode conversion of internal
waves to Alfv\'{e}n waves.

Our analysis shows that the internal waves are generated in both models and
overcome the strong radiative damping in the lower photosphere to propagate into
the higher layers. The radiative damping is strong below z=200\,km but the phase
difference spectra show signatures of these waves even below this height,
suggesting that the mechanism generating them efficiently imparts enough energy
to the wave to overcome the strong radiative damping. But the magnetic fields
affect these waves as they propagate higher up in the atmosphere as evident from
the differences between the phase difference spectra of the non-magnetic and the
magnetic model. We explore different causes that may lead to the observed
signatures and the differences in the phase difference spectra of the waves. We
conclude that the internal waves in the quiet Sun most likely undergo mode
coupling to the slow magneto-acoustic waves as described by
\citet{2010MNRAS.402..386N,
2011MNRAS.417.1162N} 
and are mostly reflected back into the atmosphere. Looking at the height
dependence of the phase spectra, we confirm that this reflection happens well
within the region where the average plasma-$\beta$ is larger than 1 (i.e. within
the gas-dominated region), confirming the mode-coupling scenario. This is also
in agreement with the energy flux spectra, which shows a mixed upward and
downward transport of energy in the internal gravity wave regime for the
magnetic case in the higher layers. Since the magnetic fields in our model are
mostly vertical, conversion to Alfv\'{e}n waves is highly unlikely. Conversion
to Alfv\'{e}n waves is not facilitated unless there is a significantly inclined
magnetic field present. The effect of the horizontal fields on the propagation
of internal waves will be explored in a later paper. We also note out that the
strong suppression that is observed within magnetic flux-concentration
\citep{2008ApJ...681L.125S} 
may be the effect of non-linear wave breaking due to the vortex flows that are
ubiquitously present in these regions. We also find that the surface-gravity
waves are strongly suppressed in the magnetic model as we go higher up in the
atmosphere, likely due to the strong vertical component of the magnetic field.

The analysis presented in this paper is based on models computed with different
numerical solvers, which resulted in a smaller size of the granules in the
non-magnetic run. However, a preliminary study using the identical MHD solver for
	both runs shows that the particular propagation properties of internal waves 
	that are found in this paper are independent of the solver. The granules are of 
	the same size and match with the sizes that we see in the magnetic model 
	of the present paper.

This analysis has shown that the internal waves are strongly affected by the
magnetic fields present on the Sun. Recognizing that a considerable amount of
internal wave flux is produced in the near surface layers, and that these waves
can couple with other magneto-atmospheric waves, it is important to fully
understand the transfer of energy from these waves to other waves in the
atmosphere of the Sun. In a broader context, a clear insight into the internal
wave spectrum will help us to connect the missing link in our understanding of
all the different wave phenomena in the solar atmosphere and their individual
role in heating the upper atmosphere either directly or indirectly.

\acknowledgements
This work was supported by a NASA EPSCoR award to NMSU under contract 
No.\,NNX09AP76A and NSF PAARE award AST-084 9986. The research leading to these
results has received funding from the European Research Council under the
European Union's Seventh Framework Programme (FP7/2007-2013) / ERC Grant
Agreement n.\,307117 and n.\,312844. 
We especially thank Stuart Jefferies for his talk at the Fifty Years of Seismology of the Sun and Stars conference in Tucson, 
that stimulated our interest in studying internal gravity waves.
The authors are grateful to Bernhard Fleck for detailed comments on a draft of this paper.
GV acknowledges the helpful discussions with
Markus Roth, Nazaret Bello Gonz{\'a}lez, Patrick Gaulme, Thierry Appourchaux, and the
{CO$^{\rm 5}$BOLD} community.
We would like to thank the anonymous referee for his/her detailed comments, which helped us to improve the paper.

\software{CO$^{\rm 5}$BOLD \citep{2012JCoPh.231..919F}}

\bibliographystyle{aasjournal}
\bibliography{vigeesh_adslinks}

\end{document}